\documentclass[aps,prl,preprint,superscriptaddress]{revtex4-2}

\usepackage{amssymb}
\usepackage{amsmath}
\usepackage{bm}
\usepackage{braket}
\usepackage{graphicx}
\usepackage{hyperref}
\usepackage{subfigure}

\begin{document}

\title{Linear Scaling Calculation of Atomic Forces and Energies with Machine Learning Local Density Matrix}

\author{Zaizhou Xin}
\affiliation{Key Laboratory of Computational Physical Sciences (Ministry of Education), Institute of Computational Physical Sciences, State Key Laboratory of Surface Physics, and Department of Physics, Fudan University, Shanghai 200433, China}

\author{Yang Zhong}
\email[]{yzhong@fudan.edu.cn}
\affiliation{Key Laboratory of Computational Physical Sciences (Ministry of Education), Institute of Computational Physical Sciences, State Key Laboratory of Surface Physics, and Department of Physics, Fudan University, Shanghai 200433, China}

\author{Xingao Gong}
\affiliation{Key Laboratory of Computational Physical Sciences (Ministry of Education), Institute of Computational Physical Sciences, State Key Laboratory of Surface Physics, and Department of Physics, Fudan University, Shanghai 200433, China}

\author{Hongjun Xiang}
\email[]{hxiang@fudan.edu.cn}
\affiliation{Key Laboratory of Computational Physical Sciences (Ministry of Education), Institute of Computational Physical Sciences, State Key Laboratory of Surface Physics, and Department of Physics, Fudan University, Shanghai 200433, China}

\date{\today}

\begin{abstract}
Accurately calculating energies and atomic forces with linear-scaling methods is a crucial approach to accelerating and improving molecular dynamics simulations.
In this paper, we introduce HamGNN-DM, a machine learning model designed to predict atomic forces and energies using local density matrices in molecular dynamics simulations.
This approach achieves efficient predictions with a time complexity of \(O(n)\), making it highly suitable for large-scale systems.
Experiments in different systems demonstrate that HamGNN-DM achieves DFT-level precision in predicting the atomic forces in different system sizes, which is vital for the molecular dynamics.
Furthermore, this method provides valuable electronic structure information throughout the dynamics and exhibits robust performance.

\end{abstract}

\maketitle

Over the past few decades, Molecular Dynamics (MD) methods have become indispensable tools in condensed matter physics, materials science, and related fields.
These computational techniques enable the simulation of atomic-scale processes, providing critical insights into the physical properties and crystal behavior of complex systems.
The advent of \textit{ab initio} Molecular Dynamics (AIMD)\cite{kresseInitioMolecularDynamics1993,solerSIESTAMethodInitio2002,hutterCPMDCarParrinelloMolecular2005,marxInitioMolecularDynamics2009}, which combines the classical MD framework with quantum mechanical calculations based on Density Functional Theory (DFT), has further advanced the field.
By enabling the direct calculation of interatomic forces, AIMD has significantly strengthened the connection between theoretical predictions and experimental observations.
However, the computational bottleneck of DFT lies in its self-consistent field (SCF) iterative procedure, which becomes increasingly demanding as system size grows.
This escalating computational cost has become a major barrier, limiting the applicability of AIMD to large-scale systems and extended timescales.

In recent years, the rapid advancement of machine learning (ML) has opened new avenues for addressing the computational challenges inherent in electronic structure calculations.
ML models trained on electronic Hamiltonian\cite{unkeSE3equivariantPredictionMolecular2021,zhangEquivariantAnalyticalMapping2022,nigamEquivariantRepresentationsMolecular2022,liDeeplearningDensityFunctional2022c,zhongTransferableEquivariantGraph2023a,zhongUniversalMachineLearning2024a} datasets have demonstrated remarkable potential for accelerating traditionally expensive steps in DFT workflows, such as the SCF iteration.
However, a significant computational hurdle remains: the diagonalization of the Hamiltonian matrix, an operation with cubic scaling $\left(O(n^3)\right)$ in the number of atoms.
This step becomes prohibitively expensive for large systems, posing a fundamental challenge to the scalability of AIMD.
Another widely adopted approach is the construction of machine learning potential (MLP) or machine learning force fields (ML-FFs)\cite{chmielaMachineLearningAccurate2017,zhangDeepPotentialMolecular2018,chmielaExactMolecularDynamics2018,wangDeePMDkitDeepLearning2018a,schuttSchNetPackDeepLearning2019,hajibabaeiUniversalMachineLearning2021,batznerE3equivariantGraphNeural2022,musaelianLearningLocalEquivariant2023,zengDeePMDkitV2Software2023,dengCHGNetPretrainedUniversal2023,qiRobustTrainingMachine2024}.
which makes it possible to study the dynamics of large-scale molecules at DFT accuracy.
However, this approach still faces challenges, notably the absence of electronic structure information and limited transferability\cite{unkeMachineLearningForce2021}.

\begin{figure*}
    \centering
    \includegraphics[width=0.9\textwidth]{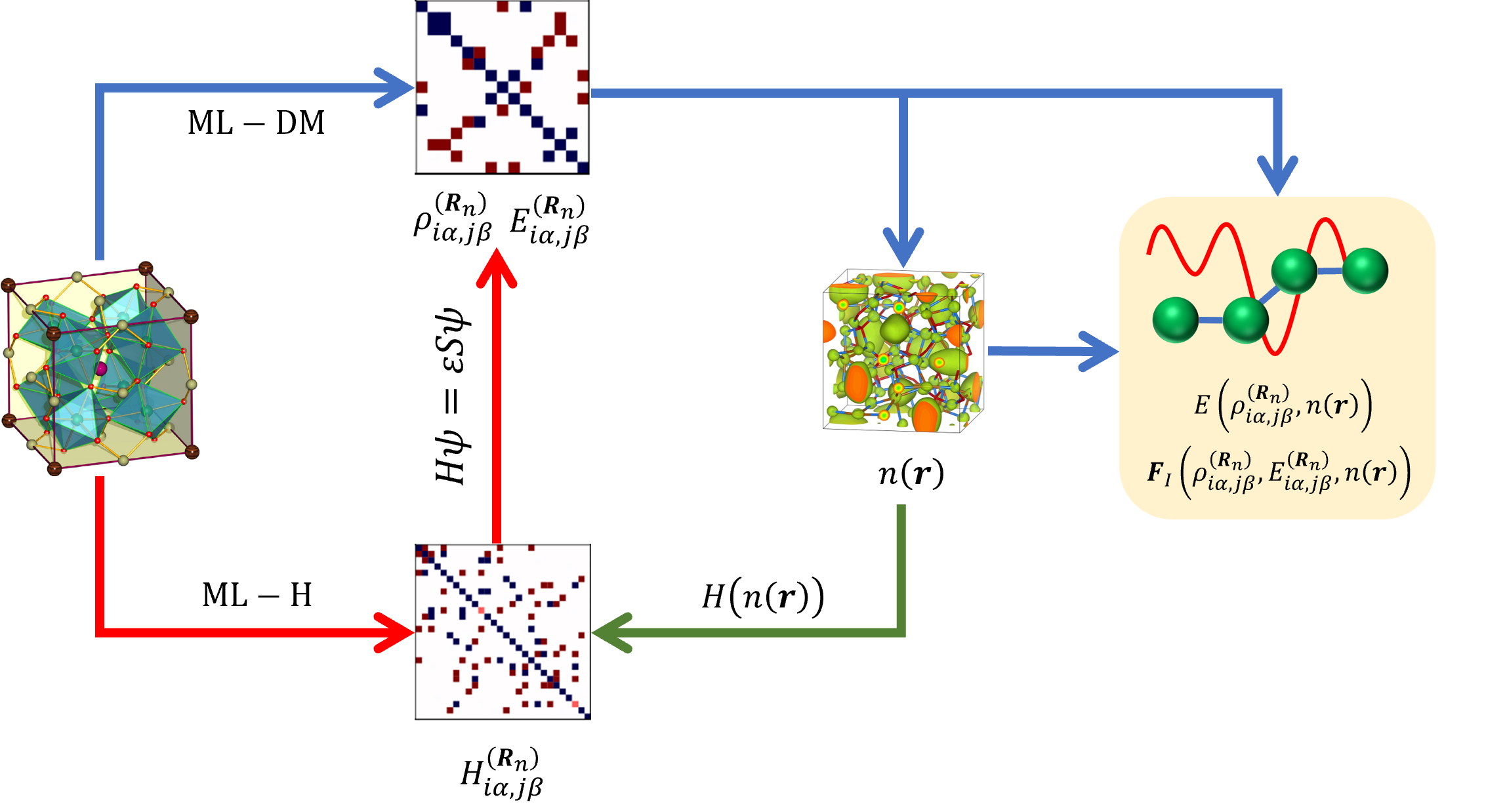}
    \caption{
    \textbf{Flowchart of the proposed method:}
    \textbf{Blue arrows:}
    Our machine learning model directly predicts the local density matrices $\rho_{i\alpha,j\beta}^{(\bm{R}_n)}$ and the local energy density matrices $E_{i\alpha,j\beta}^{(\bm{R}_n)}$ from the crystal structures.
    The charge density $n(\bm{r})$ is subsequently derived from the local density matrices.
    Using $n(\bm{r})$, $\rho_{i\alpha,j\beta}^{(\bm{R}_n)}$ and $E_{i\alpha,j\beta}^{(\bm{R}_n)}$, the system’s energy and atomic forces can be computed.
    The entire workflow of this method exhibits linear scaling with respect to system size.
    \textbf{Green arrow:}
    For cases where the Hamiltonian $H_{i\alpha,j\beta}^{(\bm{R}_n)}$ is required—such as when calculating properties like the band structure—the Hamiltonian can be constructed from the charge density $n(\bm{r})$.
    \textbf{Red arrows:}
    In contrast, when employing a machine learning Hamiltonian to compute the energy and atomic forces, the diagonalization of the Hamiltonian matrix $H_{i\alpha,j\beta}^{(\bm{R}_n)}$ is necessary.
    This diagonalization operation scales cubically with the number of atoms.
    }
    \label{fig:flowchart}
\end{figure*}

To address these limitations, this study introduces HamGNN-DM, a machine learning-based approach that enables linear-scaling \textit{ab initio} molecular dynamics simulations.
The procedure of our method is shown in Figure \ref{fig:flowchart}.
Unlike conventional methods that rely on Hamiltonian diagonalization, our approach employs an E(3)-equivariant graph neural network to directly predict the local DM and the local EDM from atomic configurations.
These quantities—central to electronic structure theory—enable the efficient computation of potential energies and atomic forces without the need for diagonalization.
By eliminating this computational bottleneck, HamGNN-DM achieves a computational complexity of $O(n)$, where the cost scales linearly with the system size.
This innovation represents a significant breakthrough in AIMD, enabling scalable simulations of large systems over extended timescales while maintaining high accuracy.
By bridging the gap between computational efficiency and predictive power, HamGNN-DM expands the applicability of AIMD to a broader range of problems in condensed matter physics and materials science, paving the way for new discoveries in these fields.

In the field of condensed matter physics, particularly in the study of solid-state material properties, solving the single-particle Schrödinger equation serves as the foundation for understanding electronic structures.
The corresponding eigenfunctions $\psi_\mu^{(\bm{k})}$ of the equation can be expanded in terms of a set of pseudo-atomic orbital basis functions $\{\phi_{i\alpha}\}$:
\begin{equation}
    \psi^{(\bm{k})}_\mu = \frac{1}{\sqrt{N}} \sum_n^N e^{\mathrm{i}\bm{R}_n \cdot \bm{k}} \sum_{i\alpha} c^{(\bm{k})}_{\mu, i\alpha} \phi_{i\alpha}\left(\bm{r} - \bm{\tau}_i - \bm{R}_n\right).
\end{equation}
The overlap matrix
\begin{equation}
    S^{(\bm{k})}_{i\alpha, j\beta} = \sum_{\bm{R}_n} e^{\mathrm{i}\bm{R}_n \cdot \bm{k}} \braket{\phi_{i\alpha}\left(\bm{r}-\bm{\tau}_i\right) | \phi_{j\beta}\left(\bm{r}-\bm{\tau}_i-\bm{R}_n\right)}
\end{equation}
represents the overlap between atomic orbitals.

The ground-state information of the entire system is fully encoded in the single-particle density matrix (DM) $\hat{\rho}$.
In the position representation $\ket{r}$, the DM can be expressed via the following integral:
\begin{equation}
    \rho\left(\bm{r},\bm{r}'\right) = \frac{1}{V_\mathrm{B}} \int_\mathrm{B} \mathrm{d}^3 \bm{k} \sum_\mu^\mathrm{occ} \braket{\bm{r} | \psi^{(\bm{k})}_\mu} \braket{\psi^{(\bm{k})}_\mu | \bm{r}'}.
\end{equation}
Here, $\mathrm{d}^3 \bm{k}$ denotes the volume element in 3-dimensional momentum space, B represents the Brillouin zone, and "occ" sums over all occupied states.
The DM decays exponentially with the increasing of $\left|\bm{r}-\bm{r}'\right|$ in semiconductors and insulators, thus is localized naturally.\cite{ismail-beigiLocalityDensityMatrix1999,maslenLocalitySparsityInitio1998}.
In metal, the locality of DM is weakened\cite{stoddartDensityMatrixMetal1977,ismail-beigiLocalityDensityMatrix1999,maslenLocalitySparsityInitio1998,goedeckerTightbindingElectronicstructureCalculations1995,goedeckerDecayPropertiesFinitetemperature1998}, but the discussion below will show that only the local part of the DM matters in the calculation of the energy and atomic forces so that the prediction of the full DM\cite{hazraPredictingOneParticleDensity2024} is not necessary.
The charge density $n(\bm{r})$, a core physical quantity in Density Functional Theory (DFT), is given by the diagonal elements of the density matrix: $n(\bm{r})=\rho(\bm{r},\bm{r})$.
The density matrix is often constructed using the Linear Combination of Atomic Orbitals (LCAO) method, where a set of pre-optimized atomic orbitals $\{\phi_{i\alpha}\}$ is chosen as the basis to achieve computational efficiency:
\begin{eqnarray}
    n(\bm{r}) =&& \sum_{\left|\bm{\tau}_i - \left(\bm{\tau}_j + \bm{R}_n\right)\right| < R_\mathrm{cut}} \sum_{i\alpha, j\beta} \rho^{(\bm{R}_n)}_{i\alpha, j\beta}\nonumber\\
    && \times \phi_{i\alpha}\left(\bm{r}-\bm{\tau}_i\right) \phi_{j\beta}\left(\bm{r}-\bm{\tau}_i-\bm{R}_n\right).
    \label{eq:charge}
\end{eqnarray}
In this equation, $R_\mathrm{cut}$ indicates the cutoff radius, while $i$ and $j$ index different atoms, and $\alpha, \beta$ label different orbitals.
The vectors $\tau_i$ and $\tau_j$ indicate the positions of atoms $i$ and $j$, and $\bm{R}_n$ is the displacement vector between periodically repeated unit cells.
The density matrix $\rho_{i\alpha,j\beta}^{(\bm{R}_n)}$ is expressed as:
\begin{equation}
    \rho_{i\alpha,j\beta}^{(\bm{R}_n)} = \frac{1}{V_\mathrm{B}} \int_\mathrm{B} \mathrm{d}^3 \bm{k} \sum_\mu^\mathrm{occ} e^{\mathrm{i}\bm{R}_n \cdot \bm{k}} c^{(\bm{k})*}_{\mu, i\alpha} c^{(\bm{k})}_{\mu, j\beta}.
\end{equation}
The subscript of the first summation in Eq.~\eqref{eq:charge} means only the part within $R_\mathrm{cut}$ is taken into account.
Due to the locality of real-space atomic orbitals, when $\left|\bm{\tau}_i - \left(\bm{\tau}_j + \bm{R}_n\right)\right| > R_\mathrm{cut}$, the overlap between two spherical harmonics becomes negligible, and their product $\phi_{i\alpha}(\bm{r} - \bm{\tau}_i) \phi_{j\beta}(\bm{r} - \bm{\tau}_j - \bm{R}_n)$ can be considered zero.
Under this approximation, only the neighboring density matrix elements $\rho_{i\alpha,j\beta}^{(\bm{R}_n)}$ contribute to the system's charge density $n(\mathrm{r})$.
Since n(r) determines the system's physical properties, these properties depend primarily on the near-neighbor terms of $\rho_{i\alpha,j\beta}^{(\bm{R}_n)}$.

Based on the density matrix and charge density, the Kohn-Sham total energy $E^{\mathrm{KS}}$ can be written as the sum of band-structure (BS) energy $E^{\mathrm{BS}}$, certain correction terms (sometimes referred to as "double-counting" corrections) $E^{\mathrm{DC}}$, and the Coulomb energy between atomic nuclei $E^{\mathrm{CC}}$\cite{solerSIESTAMethodInitio2002,garciaSiestaRecentDevelopments2020},
where
\begin{equation}
    E^{\mathrm{BS}} = \sum_{\bm{R}_n} \sum_{i\alpha, j\beta} H^{(\bm{R}_n)}_{i\alpha, j\beta} \rho^{(\bm{R}_n)}_{i\alpha, j\beta}.
\end{equation}
In the presentation of a set of pseudo-atomic orbital basis functions, the Hamiltonian $ H^{(\bm{R}_n)}_{i\alpha, j\beta}$ is localized, which vanishes when $\left|\bm{\tau}_i - \left(\bm{\tau}_j + \bm{R}_n\right)\right| > R_\mathrm{cut}$.
This indicates that $E^{\mathrm{BS}}$ only depends on the local part of the density matrix inside $R_\mathrm{cut}$.
While $E^{\mathrm{DC}}$ is a functional of the charge density $n(\bm{r})$, which relies on only the local DM, and $E^{\mathrm{DC}}$ is unrelated to the density matrix, we have proved that $E^{\mathrm{KS}}$ can be calculated from the local DM.

The force on atom $a$ can be computed as the partial derivative of the total energy with respect to the atomic coordinates.
All of the components is easily calculated from the local DM, except one term in the band-structure force component.
Alternatively, it can be transformed into another form:
\begin{equation}
    -\sum_{\bm{R}_n} \sum_{i\alpha, j\beta} H_{i\alpha, j\beta}^{(\bm{R}_n)} \frac{\partial \rho_{i\alpha, j\beta}^{(\bm{R}_n)}}{\partial \bm{\tau}_a} = \sum_{\bm{R}_n} \sum_{i\alpha, j\beta} E_{i\alpha, j\beta}^{(\bm{R}_n)} \frac{\partial S_{i\alpha, j\beta}^{(\bm{R}_n)}}{\partial \bm{\tau}_a}
\end{equation}
Here, $E_{i\alpha, j\beta}^{(\bm{R}_n)}$ is the energy density matrix (EDM), defined as:
\begin{equation}
    E_{i\alpha,j\beta}^{(\bm{R}_{n})} = \frac{1}{V_\mathrm{B}} \int_\mathrm{B} \mathrm{d}^3 \bm{k} \sum_\mu^\mathrm{occ} e^{\mathrm{i}\bm{R}_n \cdot \bm{k}} \varepsilon_{\mu}^{(\bm{k})} c_{\mu,i\alpha}^{(\bm{k})*} c_{\mu,j\beta}^{(\bm{k})}.
\end{equation}
where $\varepsilon_{\mu}^{(\bm{k})}$ are the eigenvalues of energy.
Within the DFT computational framework, evaluating the eigenvalues requires diagonalization of the Hamiltonian matrix, a process with $O(n^3)$ computational complexity, which limits the scalability of DFT for large systems.
Therefore, the energy density matrices are also predicted by the HamGNN-DM model in the workflow.
Since $\partial S_{i\alpha, j\beta}^{(\bm{R}_n)}/\partial \bm{\tau}_a$ is also localized, it suggests that we can derive atomic forces from the local density matrix and the local energy density matrix.
The remaining contributions to the force, such as $\bm{F}_a^{\mathrm{DC}} = - \partial E^{\mathrm{DC}} / \partial \bm{\tau}_i$, are functionals of the charge density   and can be computed analytically.
The interatomic Coulomb forces $\bm{F}_a^{\mathrm{CC}} = - \partial E^{\mathrm{CC}} / \partial \bm{\tau}_i$ are relatively straightforward to calculate using classical mechanics, as they do not depend on electronic structure information.

Through the incorporation of rotational symmetry constraints and locality properties, machine learning models can parametrize both the density matrix and energy density matrix, circumventing the expensive self-consistent iterations and diagonalization processes in DFT.
The parametrization of the density matrix and energy density matrix is subject to symmetry constraints.
Under the action of an $SO(3)$ rotation operator, the Hamiltonian and overlap matrices in the Schrödinger equation satisfy the equivariant relationships.
The wavefunction vector $C_\mu^{(\bm{k})}$ satisfies the equivariance condition $C_\mu^{(\bm{k})\prime} = D\left(Q\right) C_\mu^{(\bm{k})}$.
It can be proved that, under a rotational operation, the density matrix $\rho_{i\alpha,j\beta}^{(\bm{R}_n)}$ satisfies the same equivariance:
\begin{equation}
    \rho_{i\alpha,j\beta}^{(\bm{R}_n)\prime} = D\left(Q\right) \rho_{i\alpha,j\beta}^{(\bm{R}_n)} \left(D\left(Q\right)\right)^{T}
\end{equation}

The discussions on equivariance for the density matrix also apply to the energy density matrix $E_{i\alpha,j\beta}^{(\bm{R}_n)}$.
By introducing rotational symmetry constraints, the density and energy density matrices can be decomposed into spherical tensor representations that satisfy group-theoretic properties.
The property of equivariance significantly simplifies the computational complexity and facilitate the parametrization of $\rho_{i\alpha,j\beta}^{(\bm{R}_n)}$ and $E_{i\alpha,j\beta}^{(\bm{R}_n)}$ using machine learning models.
By fitting these matrices with machine learning, the expensive self-consistent iterations and diagonalization of the Hamiltonian matrix in DFT can be eliminated.

\begin{figure}
    \centering
    \subfigure{
        \includegraphics[width=0.45\textwidth]{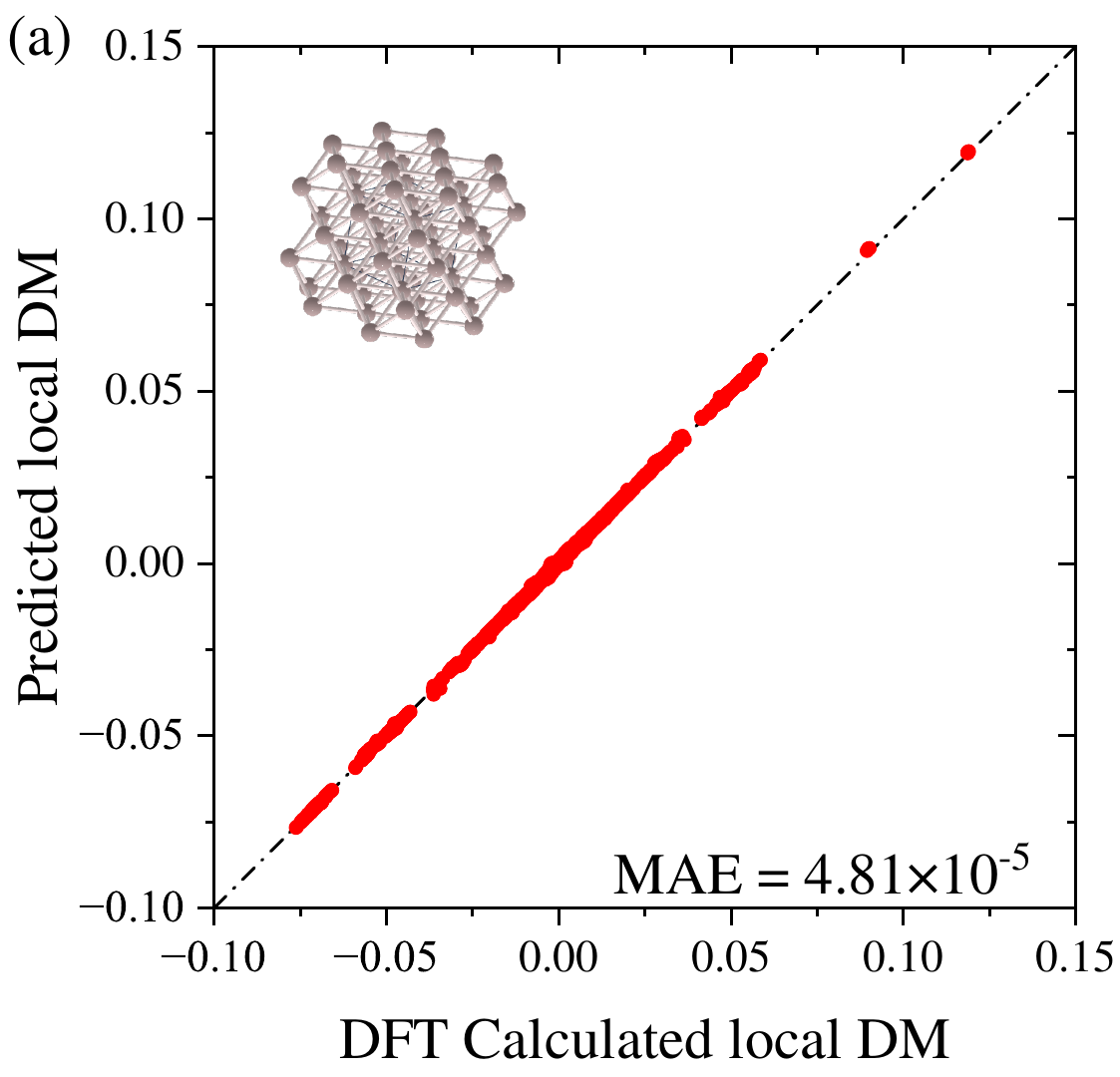}
        \label{fig:struct-al}
    }
    \subfigure{
        \includegraphics[width=0.45\textwidth]{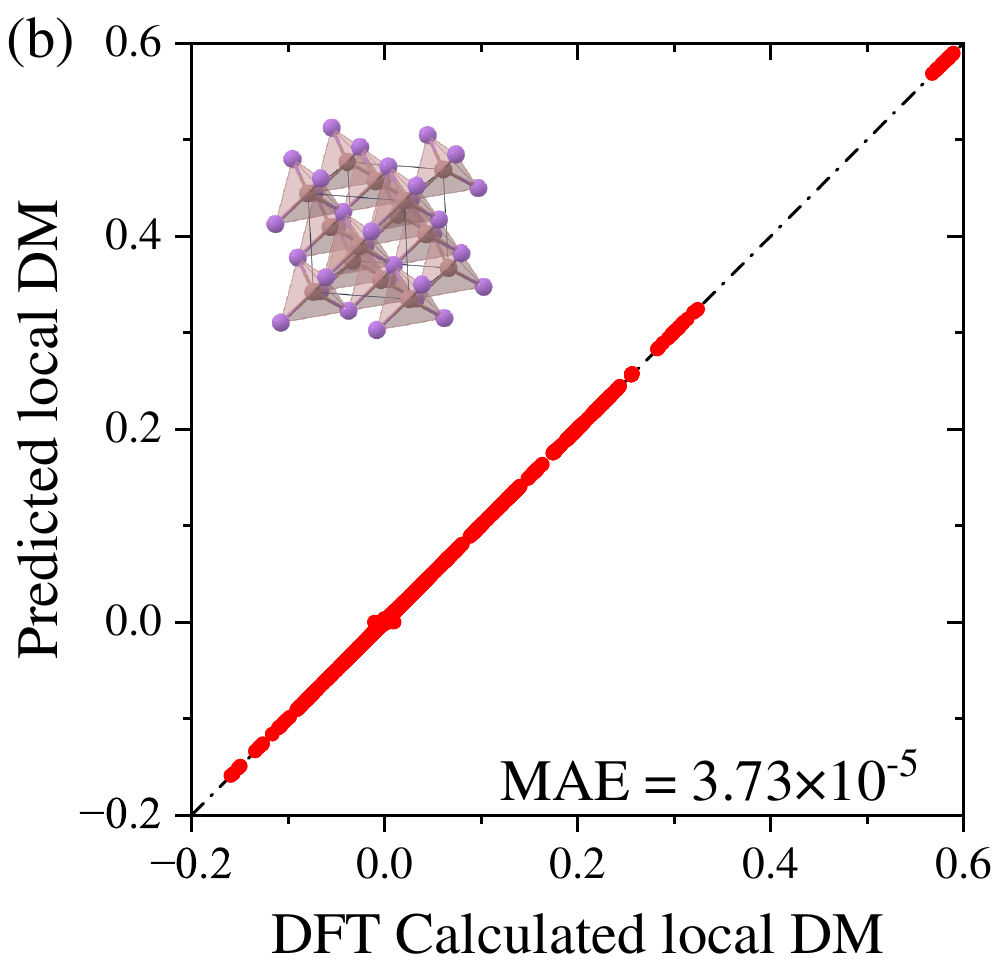}
        \label{fig:struct-GaAs}
    }
    
    \subfigure{
        \includegraphics[width=0.45\textwidth]{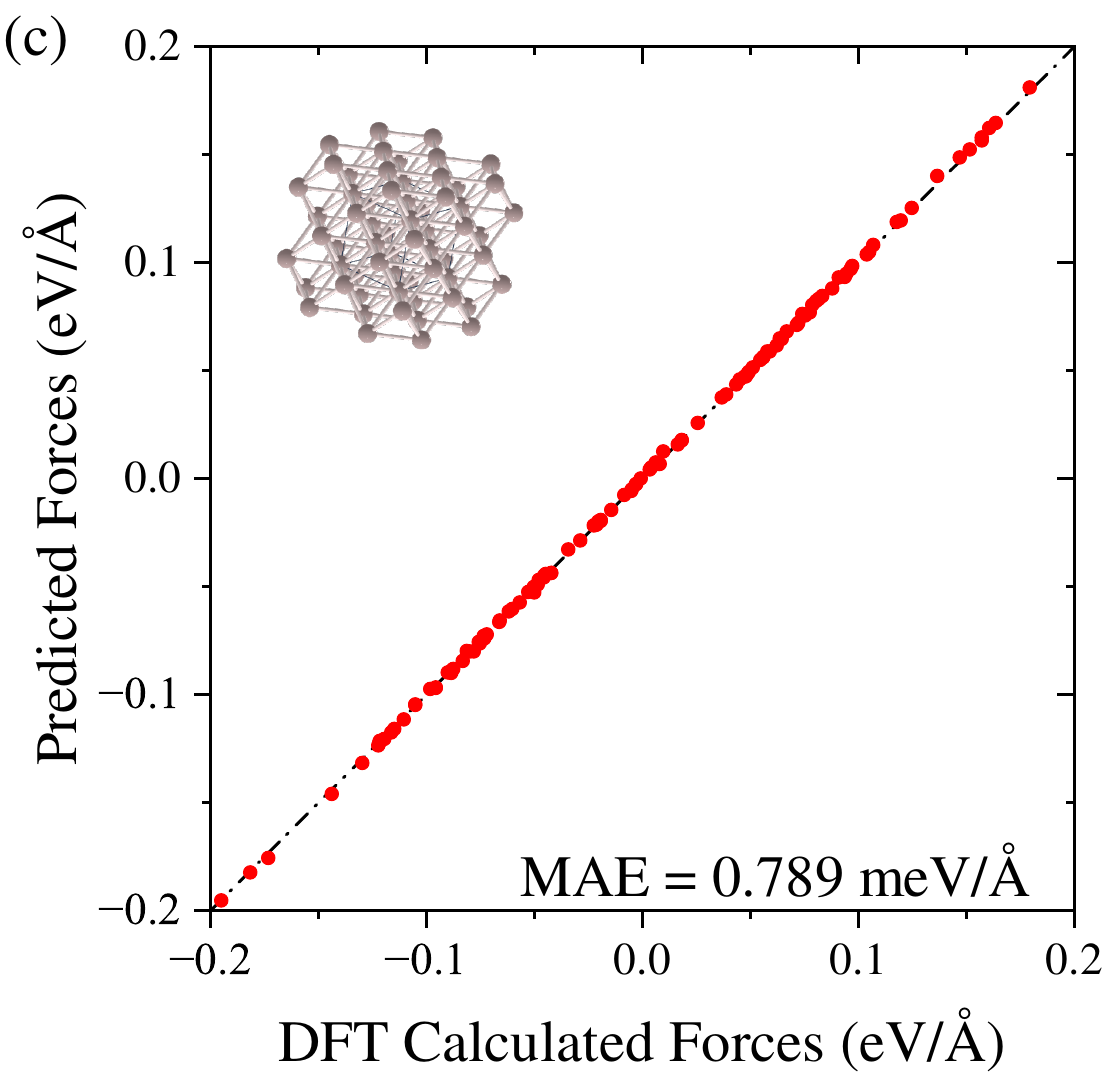}
        \label{fig:forces-al}
    }
    \subfigure{
        \includegraphics[width=0.45\textwidth]{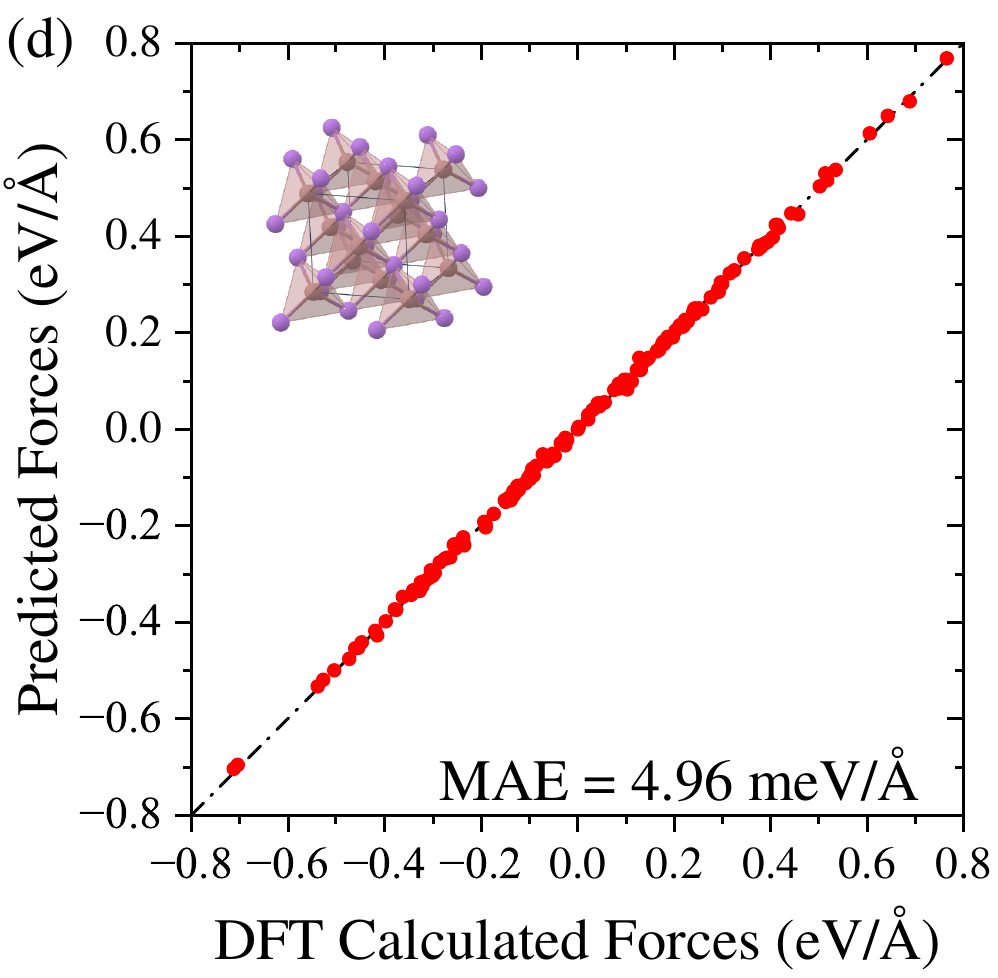}
        \label{fig:forces-GaAs}
    }
    \caption{\textbf{The prediction of HamGNN-DM on several solid systems.} \textbf{(a)-(b)} Crystal structures of Al(mp-134), GaAs(mp-2534). \textbf{(c)-(d)} Comparison of the atomic forces predicted by HamGNN-DM and those calculated by SIESTA\cite{solerSIESTAMethodInitio2002} for Al(mp-134), GaAs(mp-2534).
    The mean absolute errors of each result are shown in each figure.}
    \label{fig:comparison}
\end{figure}

To verify the effectiveness of our method, the results of some experiments will be presented.
In the first experiment, we demonstrate the accuracy of predicting the density matrix and the energy density matrix for different system, as well as the precision of the calculated atomic forces.
In the second experiment, the GaAs(mp-2534) system is expanded to investigate the computational time when the method is applied to systems of different sizes.
In experiment III, the predicted forces are used to calculate the phonon spectrum. By building Hamiltonian from the predicted DM, we can then plot the band structure and DOS of the systems.
In the final experiment, the results of our method are compared with those given by machine learning potential\cite{musaelianLearningLocalEquivariant2023}.

The HamGNN-DM model is applied to learn the density matrix and energy density matrix of multiple systems.
The reference dataset consists of 200 new structures generated by perturbing the atomic structures, with the density matrix and energy density matrix for each structure calculated by SIESTA calculations\cite{solerSIESTAMethodInitio2002,garciaSiestaRecentDevelopments2020}.
Then the trained model is applied to new structures generated via random perturbations, predicting the density matrix and energy density matrix of the new structures.

As shown in Fig. \ref{fig:comparison}, there is a high level of consistency between the predicted values of the density matrix and energy density matrix and the target values obtained from SIESTA calculations.
The mean absolute errors (MAE) of density matrix elements and energy density matrix elements are at the level of $10^{-5}$ and 0.1 $\text{meV}$ respectively, which are on the same order of tolerance set in DFT calculations.
This indicates that our model can predict the density matrix and energy density matrix with DFT-level accuracy.
Comparing the atomic forces calculated using both methods, the predicted mean absolute error is at the level of 1 \(\text{meV/\AA}\).
Additionally, we expanded the GaAs(mp-2534) system to create a 250-atom system and used the model to predict its density matrix and energy density matrix.
The mean absolute errors between the predicted values and the DFT target values are \(6.45 \times 10^{-5}\) and 0.432 \(\text{meV}\), respectively, indicating that our model maintains high predictive accuracy across systems of different sizes.
And the MAE of atomic forces is 8.98 \(\text{meV/\AA}\), still maintaining the order of magnitude of 1 \(\text{meV/\AA}\).

\begin{figure}
    \centering
    \subfigure{
        \includegraphics[width=0.45\textwidth]{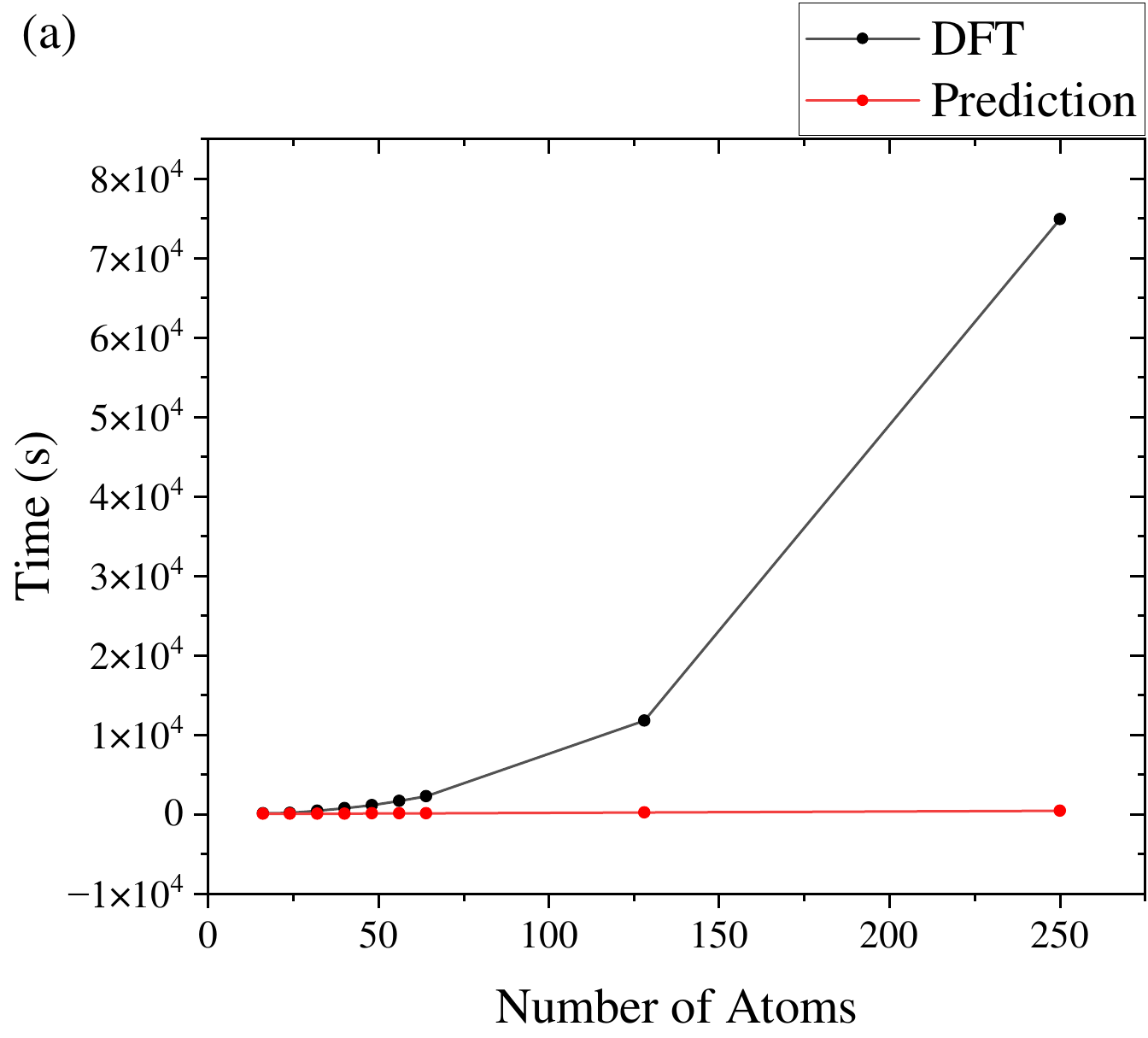}
        \label{fig:speed-test}
    }
    \subfigure{
        \includegraphics[width=0.45\textwidth]{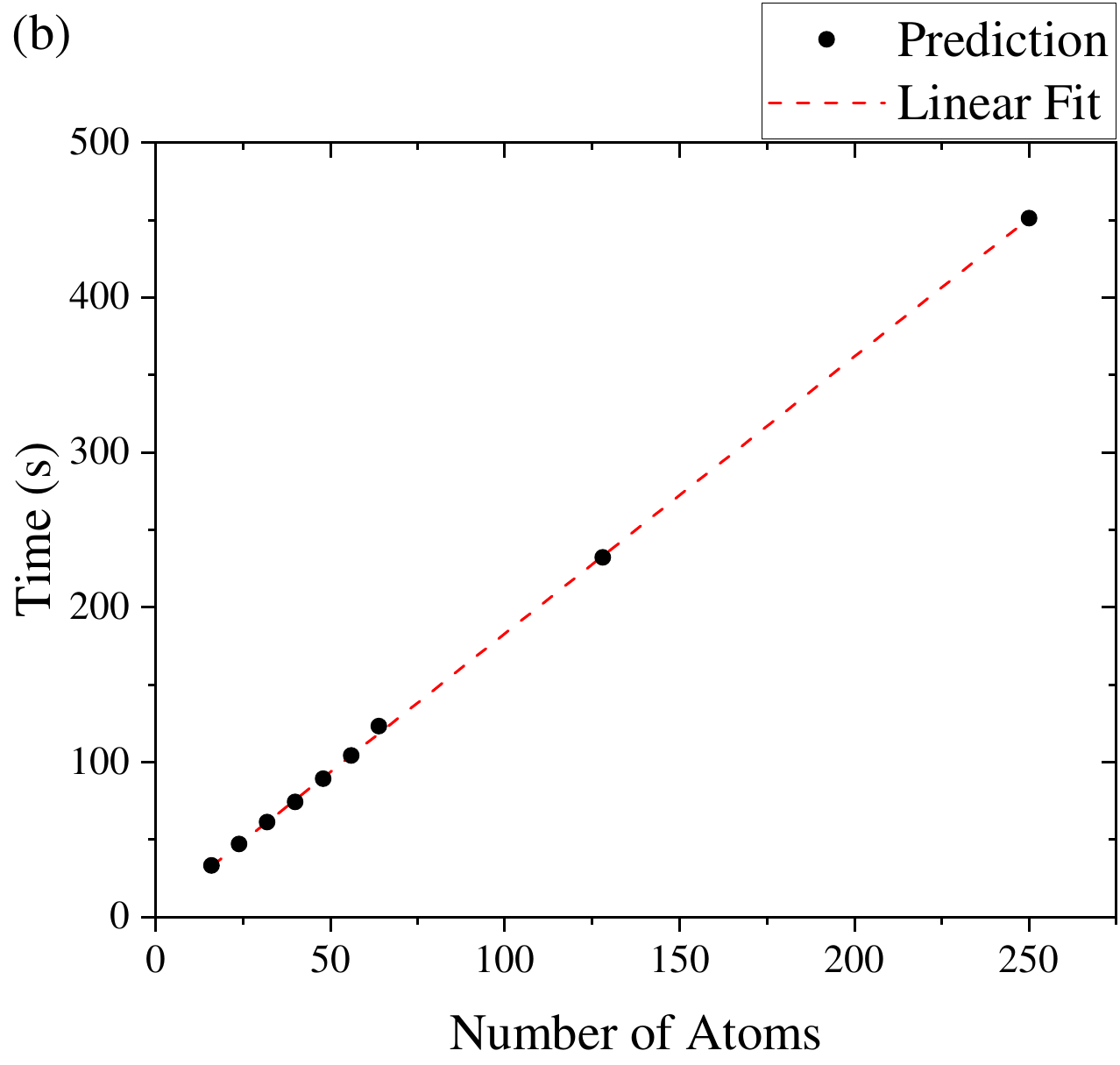}
        \label{fig:linear-fit}
    }
    \caption{\textbf{(a)} The comparison between computational time costs of the predictions by HamGNN-DM and those of the calculations by SIESTA on GaAs systems of different sizes. \textbf{(b)} The computational time costs of the prediction and their linear fit curve.}
    \label{fig:time}
\end{figure}

\begin{figure}
    \centering
    \subfigure{
        \includegraphics[width=0.45\textwidth]{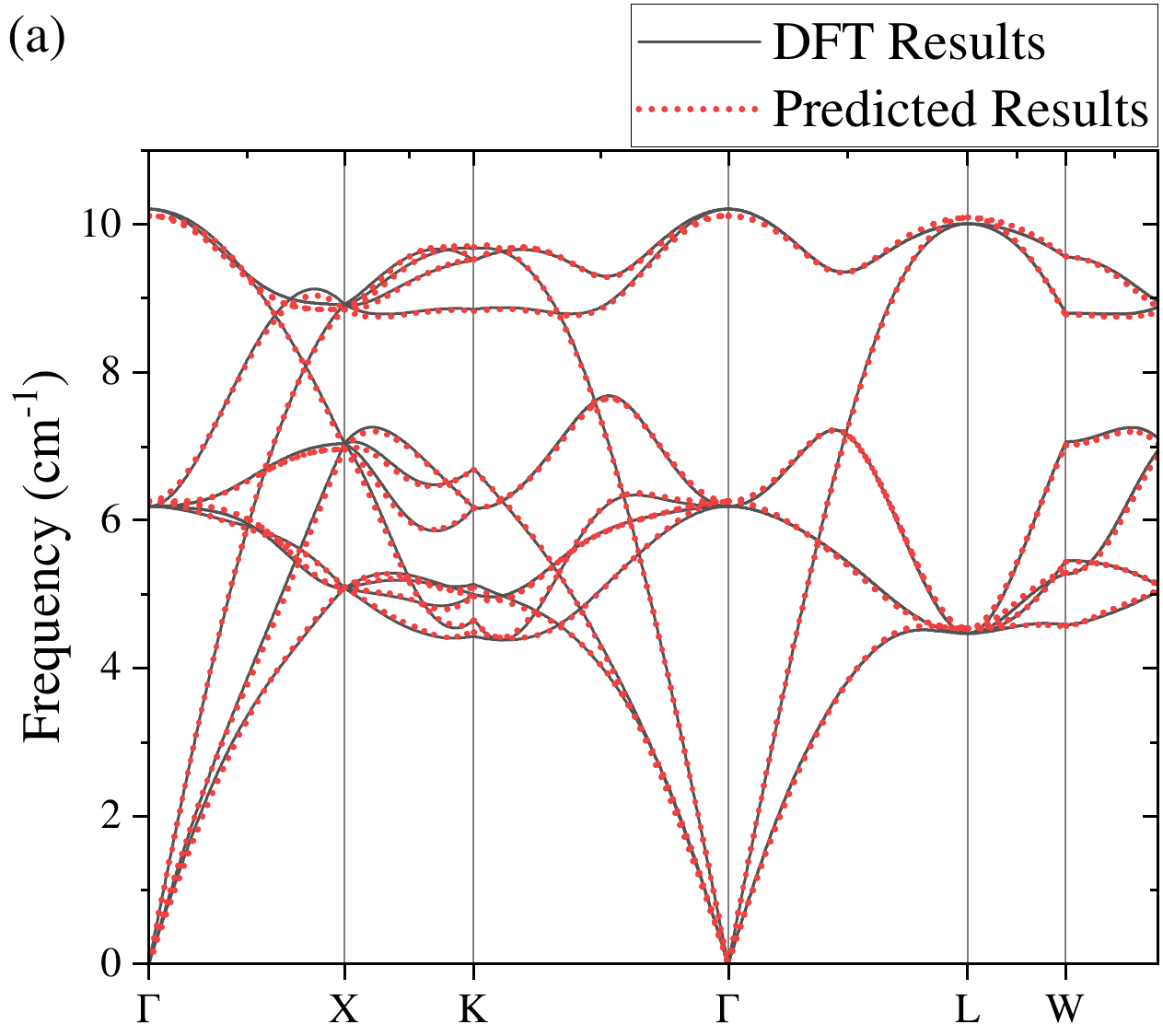}
        \label{fig:phonon-al}
    }
    \subfigure{
        \includegraphics[width=0.45\textwidth]{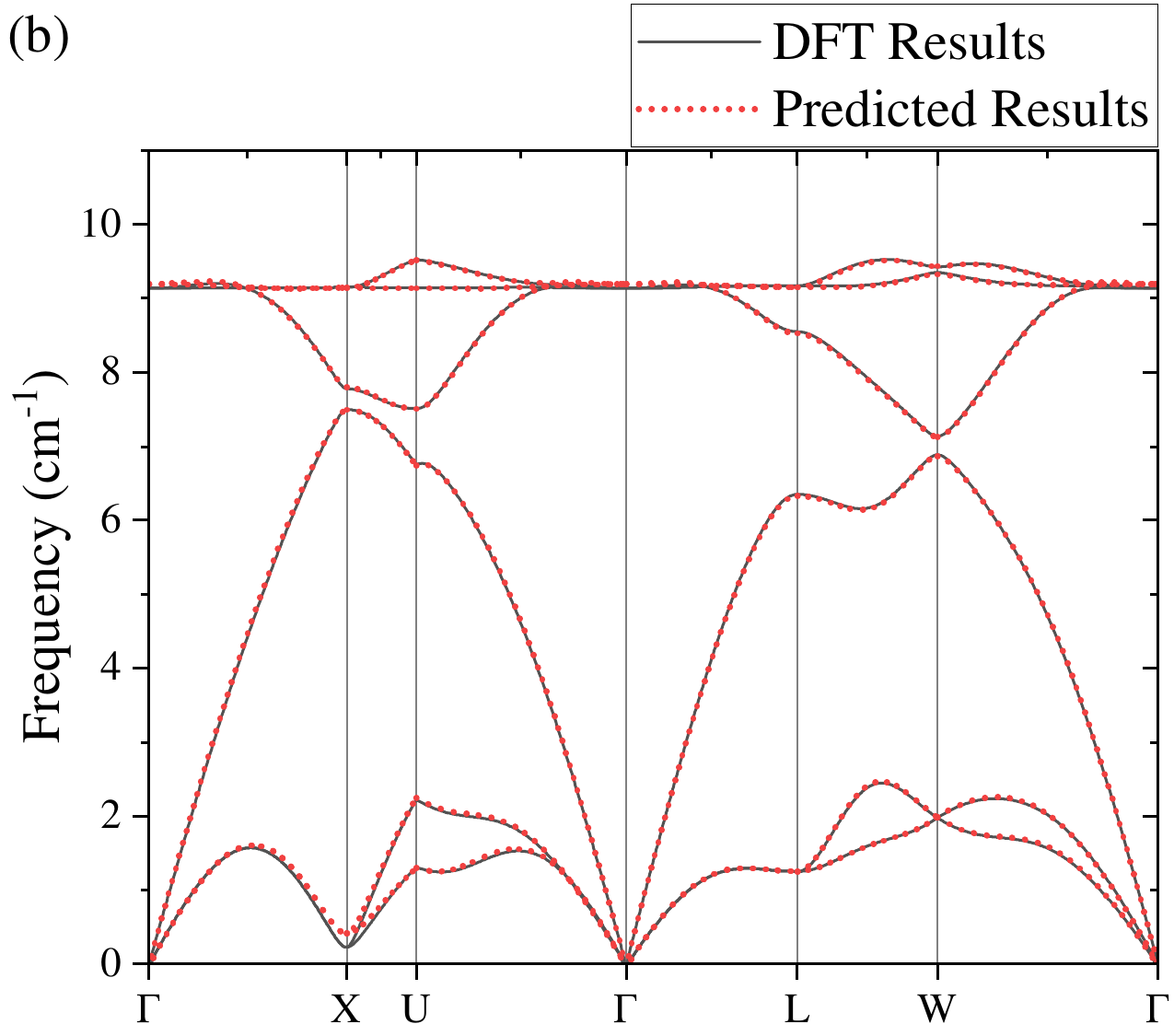}
        \label{fig:phonon-GaAs}
    }
    \caption{Comparison of phonon dispersion curves calculated by HamGNN-DM and those calculated by SIESTA for \textbf{(a)} Al, \textbf{(b)} GaAs.}
    \label{fig:phonon}
\end{figure}

\begin{figure}
    \centering
    \subfigure{
        \includegraphics[width=0.45\textwidth]{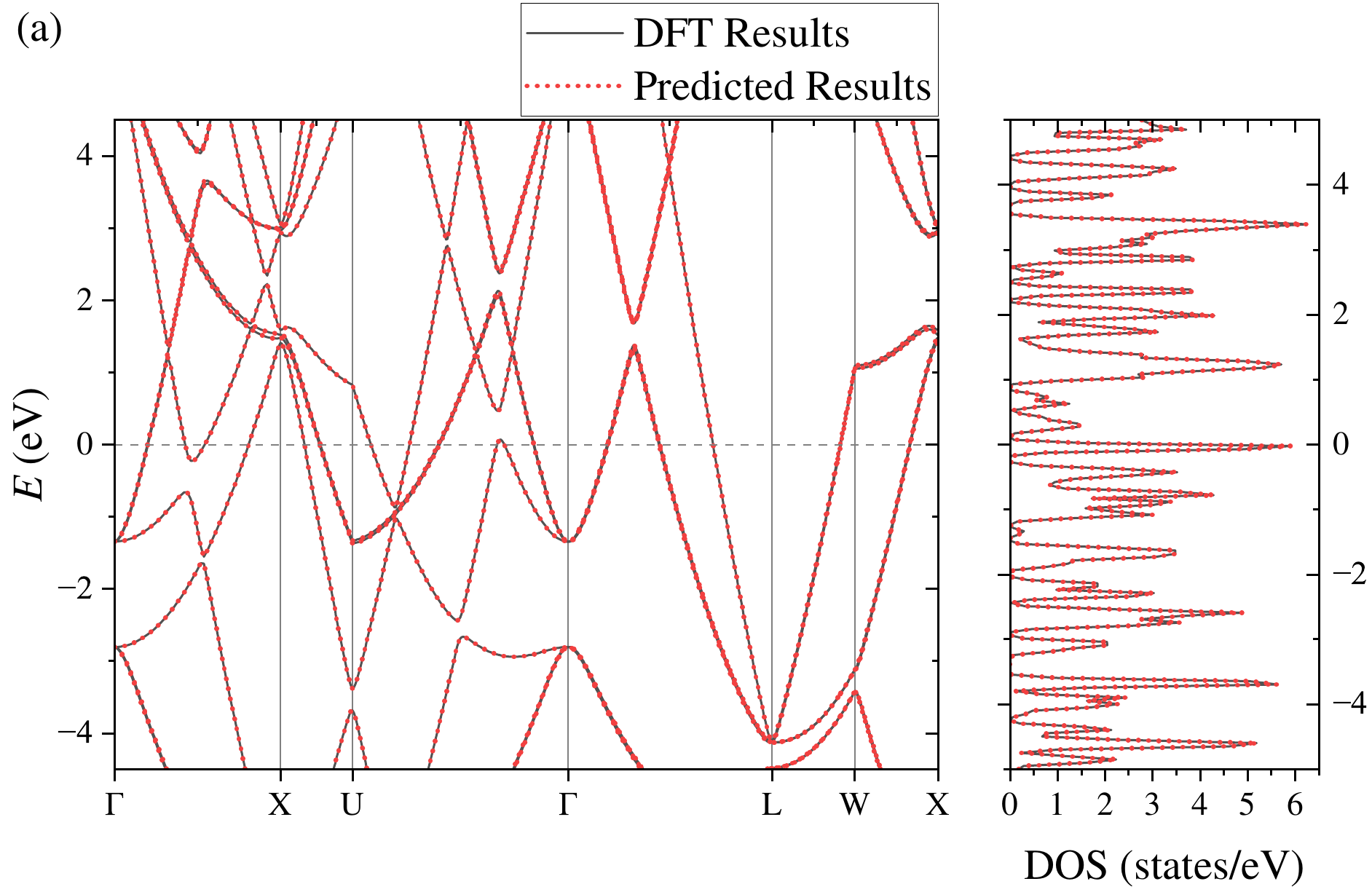}
        \label{fig:bands-al}
    }
    \subfigure{
        \includegraphics[width=0.45\textwidth]{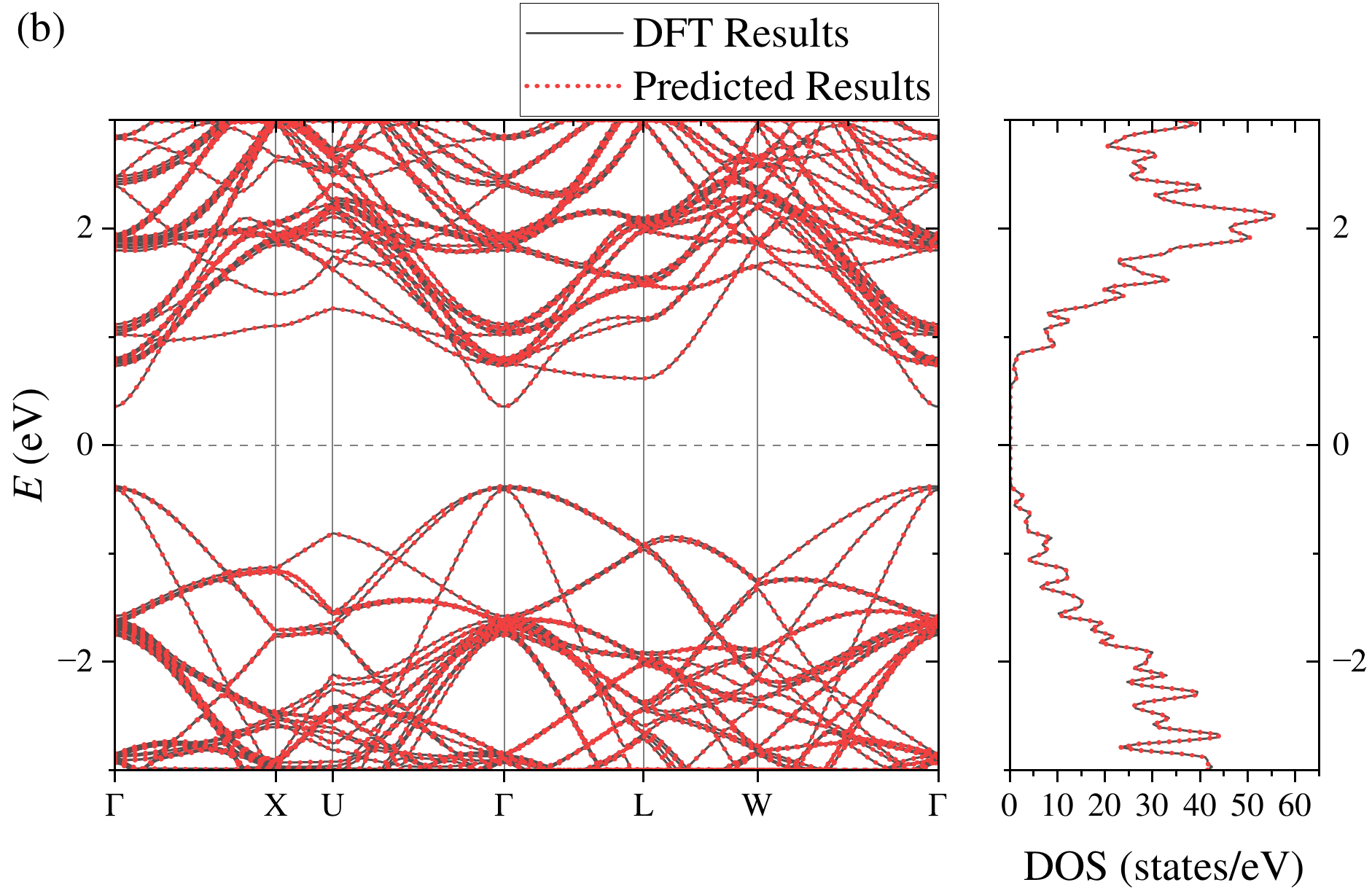}
        \label{fig:bands-GaAs}
    }
    \caption{
    Comparison of band structure and DOS calculated by HamGNN-DM and those calculated by SIESTA for \textbf{(a)} Al, \textbf{(b)} GaAs.
    To calculate the band structure, the Hamiltonian can be constructed from the local DM, and is diagonalized to get the eigenvalues of the energy.
    }
    \label{fig:bands}
\end{figure}

To test the efficiency of HamGNN-DM, the GaAs system is expanded into different scales, obtaining GaAs systems ranging from 16 to 250 atoms.
The density matrix and energy density matrix for these systems were calculated using both SIESTA and the previously trained model, and the computational time for both methods was recorded.
The results are shown in Figure \ref{fig:time}. 
As can be seen, the computational time of our method increases linearly with the system size, while the computational time of SIESTA grows cubically with the increase in system size.

\begin{figure}
    \centering
    \includegraphics[width=0.45\textwidth]{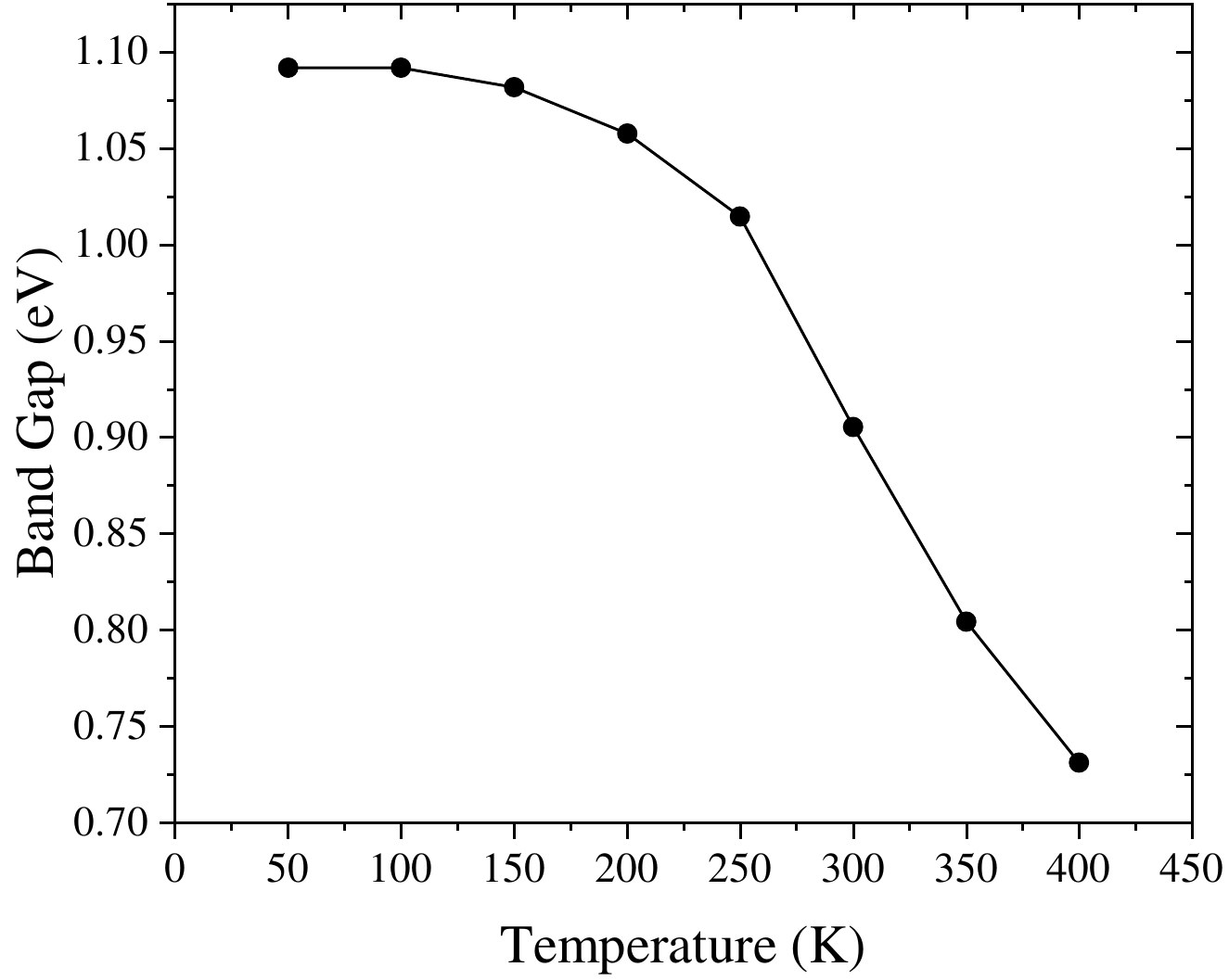}
    \caption{
    Band gap calculated by HamGNN-DM in different temperatures for GeTe.
    To calculate the band gap, the Hamiltonian can be constructed from the local DM, and is diagonalized to get the eigenvalues of the energy.
    }
    \label{fig:bandgap}
\end{figure}

After the prediction of local DM and local EDM and the calculation of energy and forces, other properties such as the phonon dispersion curves can be easily calculated\cite{togoFirstprinciplesPhononCalculations2023,togoImplementationStrategiesPhonopy2023}, shown in Figure \ref{fig:phonon}.
Besides, the Hamiltonian can also be built from DM to analyse the band structure and the DOS of the system.
This is not available in MLP, as MLP can only predict the energy and atomic forces in which the information of electronic structures is not included.
The band structures\cite{hinumaBandStructureDiagram2017} and the DOS of Al and GaAs are plotted in Figure \ref{fig:bands}.
From those results above, it is verified that the HamGNN-DM model has an excellent ability of predicting the properties including the electronic structure of both conductors and semiconductors.
Moreover, the transferability of our method, as well as that of the MLP, is also being tested.
In the previous MLP method, there is still a challenge in accurately predicting the atomic forces of carbon systems\cite{roweAccurateTransferableMachine2020,wangEnEquivariantCartesianTensor2024}.
In this test, a carbon dataset composed of 436 carbon systems with different crystal structures is prepared to train a MLP model\cite{batznerE3equivariantGraphNeural2022,musaelianLearningLocalEquivariant2023} and the model of local DM.
Then we generate 100 carbon structures by perturbing the ideal structures with position shifts smaller than 0.1 \(\text{\AA}\) to test to the prediction accuracy of the atomic forces.
The MAE of the predicted forces is 0.86 \(\text{eV/\AA}\) for our method, which behaves better than the accuracy of the MLP, 1.14 \(\text{eV/\AA}\).

Due to its incorporation of electronic structure information, the HamGNN-DM model enables the simultaneous computation of properties, such as the band gap, during molecular dynamics simulations.
For 2D mono-layer GeTe (pmmn), Nosé-Hoover molecular dynamics were performed from 400 to 50 \(\text{K}\).
At each temperature, the MD was performed for 0.2 \(\text{ns}\).
The last 10 \(\text{fs}\) was taken out to examine the change of the band gap between different temperatures.
Figure \ref{fig:bandgap} demonstrates that the band gap of GeTe decreases with increasing temperature, a behavior that is consistent with the fundamental properties of semiconductor materials\cite{capazTemperatureDependenceBand2005,giustinoElectronPhononRenormalizationDirect2010}.

In conclusion, the HamGNN-DM model demonstrates an effective and efficient approach to predicting the local density matrix, the local energy density matrix and the atomic forces for molecular dynamics simulations.
By combining the principles of equivariant machine learning with the predictive capability of graph neural networks, our model achieves DFT-level accuracy while significantly reducing computational time, especially for larger systems.
The successful application to different systems of varying sizes indicates that HamGNN-DM maintains high accuracy and efficiency across different system scales.
These results suggest that the HamGNN-DM model has a strong potential to facilitate large-scale, accurate molecular dynamics simulations, offering a promising direction for future applications in materials science, such as the calculation of the phonon spectrum and the electronic structures.

\bibliography{mlDenMat}

\end{document}